\documentclass[preprint2]{aastex631}
\begin{document}
\shorttitle{KQ Sco and UBC 1558}
\title{The Valuable Long-period Cluster Cepheid KQ Scorpii and other Calibration Candidates}
\author[0000-0001-8803-3840]{Daniel Majaess}
\affiliation{Mount Saint Vincent University, Halifax, Canada}
\email{Daniel.Majaess@msvu.ca}

\author[0000-0003-1184-1860]{David G.~Turner}
\affiliation{Saint Mary's University, Halifax, Canada}

\author[0000-0002-7064-099X]{Dante Minniti}
\affiliation{Instituto de Astrofísica, Dep.~de Ciencias Físicas, Facultad de
Ciencias Exactas, Universidad Andres Bello, Av.~Fernández Concha
700, Santiago, Chile}
\affiliation{Vatican Observatory, Specola Vaticana, V-00120, Vatican City,
Vatican City State}

\author[0000-0003-3496-3772]{Javier Alonso-Garcia}
\affiliation{Centro de Astronomía, Universidad de Antofagasta, Av.~Angamos 601, Antofagasta, Chile}
\affiliation{Millennium Institute of Astrophysics, Nuncio Monseñor
Sotero Sanz 100, Of.~104, Providencia, Santiago, Chile}

\author[0000-0001-6878-8648]{Roberto K. Saito}
\affiliation{Departamento de Física, Universidade Federal de Santa Catarina,
Trindade 88040-900, Florianópolis, Brazil}

\begin{abstract}
The classical Cepheid KQ Sco is a valuable anchor for the distance scale because of its long pulsation period ($28^{\rm d}.7$) and evidence implying membership in the open cluster UBC 1558. Analyses tied to Gaia DR3 astrometry, photometry, spectroscopy, radial velocities, and 2MASS-VVV photometry indicate a common distance of $2.15\pm0.15$ kpc (\citealt{lin21} DR3 corrections applied). Additional cluster Cepheid candidates requiring follow-up are identified, and it's suggested that a team of international researchers could maintain a cluster Cepheid database to guide the broader community to cases where consensus exists.
\end{abstract}


\section{Introduction}
There is resurgent interest in cluster Cepheids since they provide a means of (in)validating concerns regarding Planck $\Lambda$CDM $H_0$ and Gaia parallaxes \citep[e.g.,][]{ra23,wa24}\footnote{Other viewpoints include \citet{ow22} and \citet{fm23b}.}, and constrain the distance scale \citep[e.g.,][]{lin22,hao22}. The Cepheid KQ Sco ($28^{\rm d}.7$) is of particular interest granted remote extragalactic Cepheids are similarly bright long-period pulsators \citep[e.g.,][]{rie16,fm23b}.

\citet{tur79b} suggested KQ Sco ($\ell, b\simeq 340.3885,-0.7448 \degr$) may be associated with early-type stars in the broader field. Gaia observations subsequently fostered the discovery of an open cluster 6$\arcmin$ from KQ Sco \citep[UBC 1558,][]{cg22}.  \citet{lin22} did not associate KQ Sco with UBC 1558, likely owing to a parallax disparity between the cluster \citep{cg22} and Cepheid.  \citet{cg22} determined that UBC 1558 is $\simeq 3$ kpc distant (CMD result), whereas the DR3 parallax for KQ Sco formally implies $2.3$ kpc. Indeed, a revised distance for UBC 1558 presented here renders it proximate to the Cepheid.

In this study, DR3 spectroscopy, astrometry, photometry, and radial velocities are inspected in tandem with 2MASS-VVV photometry to clarify the connection between KQ Sco and UBC 1558 (Fig.~\ref{fig-fov}).  Lastly, a list of reputed cluster Cepheids is provided which would likewise benefit from further examination on a case-by-case basis.

\section{Analysis}
\begin{figure}[t]
\begin{center}
 \includegraphics[width=3.4in]{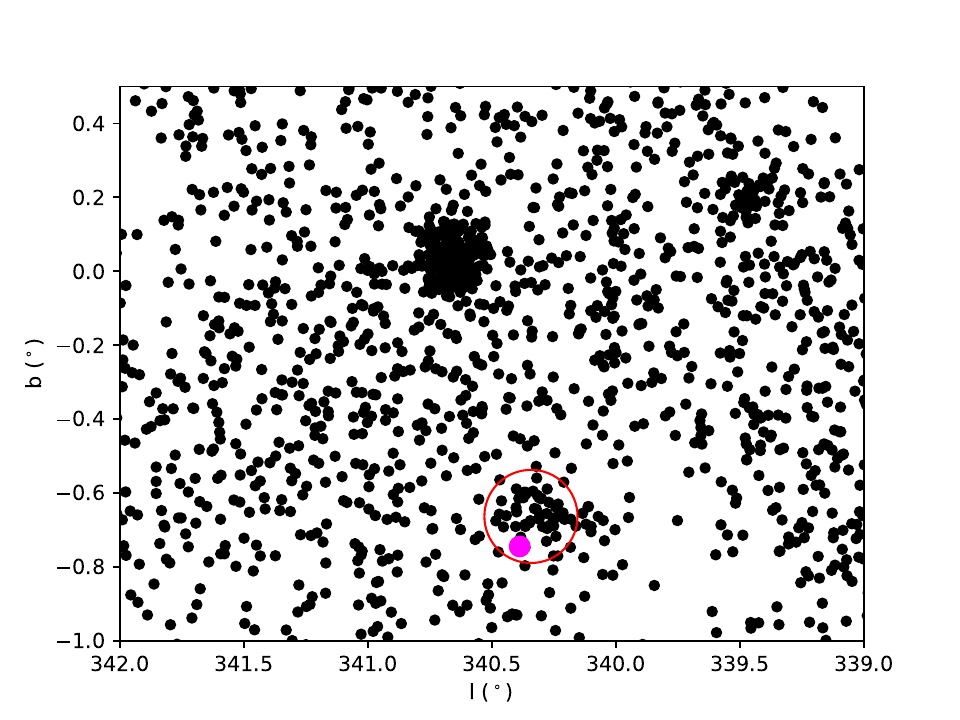} 
  \caption{Stars featuring DR3 proper motions similar to the classical Cepheid KQ Sco (magenta). The open cluster UBC 1558 \citep{cg22} is encompassed by a red circle.}
 \label{fig-fov}
\end{center}
\end{figure}

Fig.~\ref{fig-fov} highlights stars with proper motions ($\mu_{\alpha}$, $\mu_{\delta}$) within $0.2$ mas yr$^{-1}$ of KQ Sco \citep{gaia23}.  An overdensity is readily discernible which represents UBC 1558, and KQ Sco is within the confines of the cluster (Fig.~\ref{fig-fov}). UBC 1558 is not densely populated, and may not remain bound as the average cluster dissolution timescale is surpassed \citep[10 Myr,][]{bon11}.  The Gaia DR3 astrometric solutions for the cluster and KQ Sco are summarized in Table~\ref{table:summary}.  The parallaxes agree to within the uncertainties. A Wesenheit function (Leavitt Law) was employed to independently assess the distance to KQ Sco.  The compilation of \citet{ng12} presents a Cepheid Wesenheit distance of $\simeq 2.2$ kpc, and the relation of \citet{maj13} corroborates the result.  Those distances are nearer, thereby hinting at a correction to the DR3 results \citep[e.g.,][]{lin21,ow22}. Applying the \citet{lin21} prescription bridges the gap, and suggests a mean systematic correction of $\Delta \pi \simeq0.03$ mas.  For example, KQ Sco shifts from 0.43 to 0.47 mas, which is a $\simeq 10$\% distance offset (pc).

\begin{deluxetable}{l|cc}
\tablecaption{KQ Sco \& UBC 1558 Parameters.\label{table:summary}}
\tablehead{\colhead{} & \colhead{KQ Sco} & \colhead{UBC 1558}}
\startdata
$\pi_{DR3}$ (mas) & $0.43\pm0.02$ & $0.41\pm0.03$  \\
$\pi_{DR3-L21}$ (mas) & $0.47$ & $0.45$  \\
$\mu_{\alpha}$ (mas/yr) & $-1.37\pm0.02$ & $-1.34\pm0.01$  \\
$\mu_{\delta}$ (mas/yr) & $-2.50\pm0.02$ & $-2.49\pm0.01$  \\
$\log{\tau}$ & $7.48\pm0.15$ & $7.55\pm0.10$ \\
$d_{W}$ (mas) & $0.45$ & $-$ \\
$d_{J-JH}$ (mas) & $-$ & $0.46^{+0.06}_{-0.03}$ \\
\enddata
\tablenotetext{ }{* Uncertainties for cluster astrometry are the standard error. $\pi_{DR3-L21}$ represents the corrected parallax following \citet{lin21}. $d_{W}$ is the Wesenheit distance, and $d_{J-JH}$ represents an isochrone fit to 2MASS-VVV photometry.}
\end{deluxetable}

\begin{figure}[t]
\begin{center}
 \includegraphics[width=3.4in]{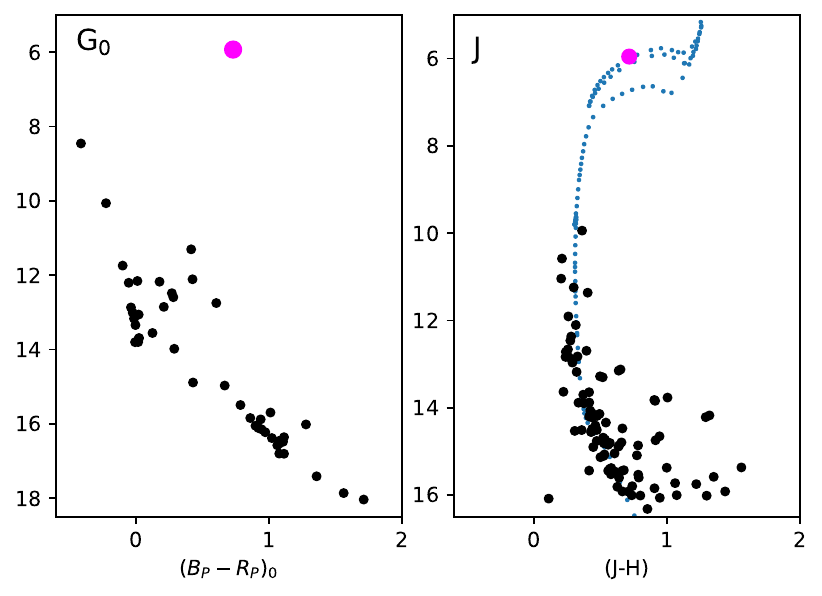} 
  \caption{Left, dereddened $G_0/(B-R)_0$ color-magnitude diagram for UBC 1558.  The main-sequence and turnoff morphology are conducive to a younger open cluster that could host a longer-period $28^{d}.7$ classical Cepheid. Right, 2MASS-VVV near-infrared $JH$ color-magnitude diagram for UBC 1558.  KQ Sco ($\log{\tau}=7.48\pm0.15$, Cepheid period-age relation) is conveyed by the magenta datum, and the $\log{\tau}=7.55$ isochrone is the dotted line.}
 \label{fig-JH}
 \label{fig-CMDs}
\end{center}
\end{figure}

The Cepheid and cluster are in temporal agreement (Table~\ref{table:summary}).  The age for KQ Sco was determined using relations derived by \citet{bo05}, \citet{tu12}, and \citet{and16}.  The estimated Cepheid age is $\log{\tau}=7.48\pm0.15$.  \citet{cg22} determined $\log{\tau}=7.25$ for UBC 1558, whereas \citet{cav24} favored $\log{\tau} \gtrsim 7.9$, and thus a redetermination of the cluster age was pursued below that yields $\log{\tau}=7.55\pm0.10$. The $B_PGR_P$ differentially dereddened color-magnitude diagram indicates that UBC 1558 is a younger open cluster (Fig.~\ref{fig-CMDs}), and was constructed using preliminary DR3 spectroscopically determined data such as $A_G$ and $E(B_P-R_P)$ \citep{gaia23}. However, there are caveats linked to that advantageous new dataset, and research is ongoing to rectify problems therein \citep{an23}.  For example, \citet{mt24} noted that certain unobscured clusters do not align with DR3 spectroscopically dereddened observations, and hence currently, $T_{\rm eff}-(B_P-R_P)_0$ inhomogeneities in the preliminary DR3 spectroscopic data inhibit a reliable isochrone fit in dereddened space.  Consequently, DR3 astrometrically cleaned near-infrared 2MASS-VVV observations were analyzed \citep{cut03,min11,sai12}, and the VVV data were standardized to 2MASS using common stars in the field.  Furthermore, reddening and extinction law variations and their uncertainties are smaller in the infrared \citep{maj16}. \citet{sa00} scaled-solar isochrones were applied to the 2MASS-VVV photometry for UBC 1558 \citep[see also][]{bon04}. The isochrone was shifted using a mean reddening inferred from a suite of B-stars confirmed by DR3 spectroscopy.  The intrinsic near-infrared colors of \citet{sl09} were utilized, and yielded a cluster reddening of $E(J-H)=0.39\pm0.03$. The ensuing cluster parameters are $d=2.1\pm0.2$ kpc and $\log{\tau}=7.55\pm0.10$ (Fig.~\ref{fig-JH}).  JH data for KQ Sco were drawn from the \citet{br21} compilation \citep{we84,ls92}.

\begin{figure}[t]
\begin{center}
 \includegraphics[width=3.5in]{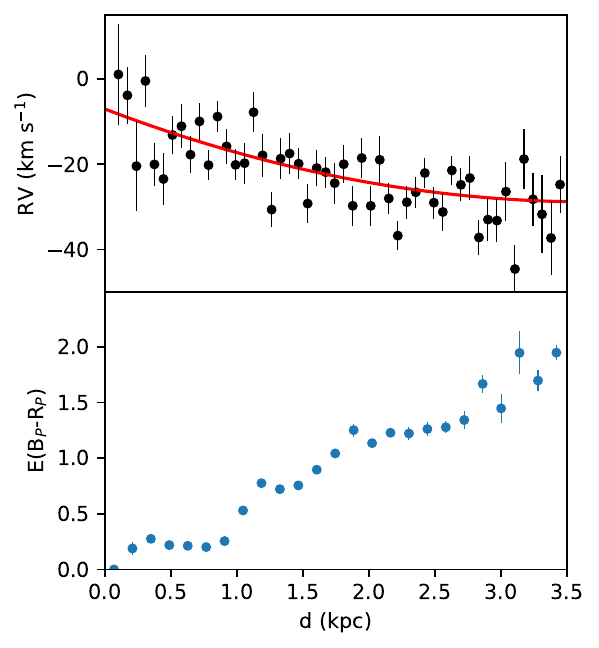} 
  \caption{Top, the binned distribution of radial velocities indicates the classical Cepheid KQ Sco ($-22.1\pm1$ km s$^{-1}$) lies near $\simeq 2$ kpc. The red line represents a polynomial fit ($2^{nd}$ order).  Bottom, UBC 1558 may reside in the interarm region between the Carina (reddening increase near $\simeq 1$ kpc) and Centaurus-Norma ($\gtrsim 2.8$ kpc) spiral arms, according to the map of \citet{Xu23}. Uncertainties in both panels reflect the standard error.}
 \label{fig-RV-red}
\end{center}
\end{figure}

The radial velocity trend along the sightline was investigated by inspecting existing and new velocities for KQ Sco \citep{cc85,and24}, and DR3 measurements.  The two cited studies relay a velocity for KQ Sco of $-22.1\pm1$ and $-23.777\pm0.052$ km s$^{-1}$, accordingly. Fig.~\ref{fig-RV-red} (top panel) conveys stars within $30\arcmin$ of KQ Sco featuring DR3 radial velocities, and adhering to canonical culling criteria (e.g., RUWE$<2$). The mean trend over the baseline shown ($2^{nd}$ order polynomial) indicates the Cepheid is $\simeq 2$ kpc away.  The originally cited $\simeq 3$ kpc distance for UBC 1558 instead points to $\simeq -29$ km s$^{-1}$.  

Fig.~\ref{fig-RV-red} (bottom panel) relays the trend of distance with reddening along the KQ Sco sightline.  Step-functions can be indicative of interstellar clouds or traversal across a spiral arm.  A comparatively nearby cloud is responsible for the initial color-excess near 0.2 kpc, and thereafter the Carina arm is traversed at this longitude ($\ell\simeq 345$ $^{\circ}$, $d\simeq 1$ kpc). A subsequent slightly positive linear trend is apparent from $\simeq 2-2.8$ kpc, whereafter the Centaurus\footnote{\citet{maj09} and \citet{Xu23} advocate that a perfect spiral pattern does not characterize the Milky Way, and there is evidence of branching (e.g., Centaurus).}-Norma spiral arm emerges according to the \citet{Xu23} map. UBC 1558 may inhabit the interarm region between the Carina and Centaurus-Norma spiral arms, yet note that Galactic structure is contested.

\begin{deluxetable*}{lccccc}
\tablecaption{Understudied cluster Cepheid candidates warranting follow-up.\label{table:cep}}
\tablehead{\colhead{ID} & \colhead{Mode} & \colhead{P (d)}&\colhead{ID} & \colhead{Mode} & \colhead{P (d)}}
\startdata
AN Aur	&	F	&	10.3	&	X Pup	&	F	&	26.0	\\
AP Cas	&	F	&	6.8	&	V335 Pup	&	1O	&	4.9	\\
BB Cen	&	1O	&	4.0	&	V724 Pup	&	F	&	5.6	\\
BV Cas	&	F	&	5.4	&	ATO	J297.7863+25.3136	&	1O	&	2.9	\\
CD Cyg	&	F	&	17.1	&	ATO	J300.0102+29.1869	&	F	&	18.4	\\
CM Sct	&	F	&	3.9	&	ZTF J192152.00+150346.9	&	F	&	8.8	\\
CN Sct	&	F	&	10.0	&	ASAS J075840-3330.2	&	F	&	4.4	\\
CS Vel	&	F	&	5.9	&	ASAS J115701-6218.7	&	F	&	26.5	\\
CV Mon	&	F	&	5.4	&	ASAS J183904-1049.3	&	1O	&	3.1	\\
DP Vel	&	F	&	5.5	&	ASASSN-V J040516.13+555512.9	&	1O	&	1.8	\\
EX Vel	&	F	&	13.2	&	ASASSN-V J151832.37-580128.7	&	F	&	9.2	\\
FF Car	&	F	&	16.3	&	ASASSN-V J194806.54+260526.1	&	1O	&	6.6	\\
FM Car	&	F	&	7.6	&	ASASSN-V J201151.18+342447.2	&	F	&	9.8	\\
FZ Car	&	1O	&	3.6	&	ASASSN-V J211659.94+514556.7	&	F	&	5.9	\\
GI Cyg	&	F	&	5.8	&	ASASSN-V J213533.70+533049.3	&	1O	&	3.2	\\
GQ Vul	&	F	&	12.7	&	NSVS 11232104	&	F	&	6.9	\\
IM Car	&	F	&	5.3	&	OGLE GD-CEP-0422	&	F	&	4.6	\\
NO Cas	&	1O	&	2.6	&	OGLE GD-CEP-0549	&	F1O	&	2.2/1.6	\\
OO Cen	&	F	&	12.9	&	OGLE GD-CEP-0605	&	1O	&	3.1	\\
RS Ori	&	F	&	7.6	&	OGLE GD-CEP-0609	&	F	&	25.1	\\
SV Cru	&	F	&	7.0	&	OGLE GD-CEP-1544	&	F	&	5.5	\\
SV Vul	&	F	&	44.9	&	Gaia DR3 5254518760118884864	&	1O	&	3.8	\\
TY Sct	&	F	&	11.1	&	Gaia DR3 5878427527969505024	&	1O	&	0.3	\\
X Cru	&	F	&	6.2	&	Gaia DR3 5935070926723295232	&	F	&	15.3	\\
\enddata
\tablenotetext{ }{Notes: IDs, pulsation modes and periods, mainly from the \citet{pie21} catalog. Candidates can overlap with other published tables, for example, DP Vel is ruled out as a cluster member by \citet{lin22}, whereas \citet{hao22} include it in a Class B subsample, and it is absent from \citet{ra23}. Hence the impetus for a cluster Cepheid database site.}
\end{deluxetable*}

\section{Conclusions}
KQ Sco exhibits parameters which are consistent with UBC 1558, and the Cepheid is a probable member (Fig.~\ref{fig-fov}, Table~\ref{table:summary}). The distance to UBC 1558 was revised nearer by $\simeq -0.8$ kpc. An unweighted mean and standard deviation associated with all distance methods yield $d=2.15\pm0.15$ kpc (excluding uncorrected DR3 parallaxes).  A separate investigation is desirable to examine whether bound or dissolving clusters encompassing UBC 1558 (Fig.~\ref{fig-fov}) are associated (e.g., Ruprecht 121).  For example, the denser open cluster NGC 6216 may be in relative vicinity to UBC 1558, yet the former may be potentially older and therefore unrelated.

Going forward, numerous understudied potential Cepheids may be cluster members (Table~\ref{table:cep}). The targets were identified by cross-referencing the \citet{pie21} Cepheid catalog with DR3.  A subsample were discussed previously \citep[e.g.,][]{zc21,lin22}, and all candidates may benefit from individual follow-up, which presents pertinent insights as conveyed here.  

More broadly, owing to the importance attributed to cluster Cepheids in diverse endeavors (e.g., ascertaining whether the Planck $\Lambda$CDM $H_0$ requires adjustment): a database site could be constructed to provide \textit{suggested} guidance for the broader community whereupon (I), an updated list of cluster Cepheids unanimously agreed upon by a panel of international researchers is presented, and (II), reputed cluster Cepheids lacking consensus are highlighted which may require additional funding to secure data and undertake a comprehensive analysis.  For example, QZ Nor \citep{eg83,maj13d} is featured in the Gold sample of \citet{ra23} \citep[see also][]{bre20}, but absent from \citet{med21} and \citet{hao22}.  Conversely, GQ Vul is classified within the Class A sample of \citet{hao22}, and is absent from \citet{ra23}, and a 0.47 membership probability was assigned by \citet[][their Table 1]{med21}.  \citet{tu08} suggest CG Cas is a member of Berkeley 58, whereas \citet{wa24} indicate it is a member of NGC 7790.  Therein lies the motivation for researchers to maintain a cluster Cepheid site.  

\begin{acknowledgments}
\small{\textbf{Acknowledgments}: this research relies on initiatives such as CDS, NASA ADS, arXiv, Gaia, Hipparcos, VELOCE, OGLE, 2MASS, and VVV (data from the ESO Public Survey program IDs 179.B-2002 and 198.B-2004, taken with the VISTA telescope, and data products from the Cambridge Astronomical Survey Unit). Dante M.~gratefully acknowledges support from the Center for Astrophysics and Associated Technologies (CATA), the ANID BASAL projects ACE210002 and FB210003, and Fondecyt Project No.~1220724.  R.~K.~S.~acknowledges support from CNPq/Brazil through projects 308298/2022-5 and 421034/2023-8.}
\end{acknowledgments}

\bibliography{article}{}
\bibliographystyle{aasjournal}
\end{document}